\documentclass[apj]{emulateapj}

\newcommand{\fdegs}{\hbox{$^{\circ}$}}

\newcommand{\papone}{Paper\,\textsc{I}}
\newcommand{\Oiii}{[O\,\textsc{iii}]}
\newcommand{\Oii}{[O\,\textsc{ii}]}
\newcommand{\Oi}{[O\,\textsc{i}]}
\newcommand{\Nii}{[N\,\textsc{ii}]}
\newcommand{\Sii}{[S\,\textsc{ii}]}

\newcommand{\Ariii}{[Ar\,\textsc{iii}]}
\newcommand{\Ha}{H$\alpha$}

\newcommand{\Hb}{H$\beta$}

\newcommand{\Hii}{H\,\textsc{ii}}
\newcommand{\Hei}{He\,\textsc{i}}
\newcommand{\Heii}{He\,\textsc{ii}}

\newcommand{\Neiii}{[Ne\,\textsc{iii}]}
\newcommand{\Nev}{[Ne\,\textsc{v}]}

\slugcomment{ApJL publication}
\shorttitle{Ionization Cone in LMC\,X-1}
\shortauthors{Cooke et al.}

\begin{document}

\title{Ionization Cone in the X-ray Binary LMC\,X-1}

\author{R. Cooke\altaffilmark{1}, J. Bland-Hawthorn\altaffilmark{1}, R. Sharp\altaffilmark{2}, Z. Kuncic\altaffilmark{1}}
\altaffiltext{1}{School of Physics, University of Sydney, NSW 2006, Australia}
\altaffiltext{2}{Anglo-Australian Observatory, Epping, NSW 1710, Australia}
\email{r.cooke@physics.usyd.edu.au}

\begin{abstract}

In an earlier paper, we presented the first evidence for a bow-shock nebula surrounding the X-ray binary LMC\,X-1 on a scale of $\sim 15$ pc, which we argued was powered by a jet associated with an accretion disk. We now present the first evidence for an ionization cone extending from an X-ray binary, a phenomenon only seen to date in active galactic nuclei (AGN). The ionization cone, detected in the \Heii$\lambda4686$/\Hb\ and \Oiii$\lambda5007$/\Hb\ line ratio maps, aligns with the direction of the jet inferred from the bow-shock nebula. The cone has an opening angle $\approx45\fdegs$ and radial extent $\approx 3.8 \, {\rm pc}$. Since the \Heii\ emission cannot be explained by the companion O star, the gas in the ionization cone must be exposed to the `naked' accretion disk, thereby allowing us to place constraints on the unobservable ionizing spectrum. The energetics of the ionization cone give unambiguous evidence for an ``ultraviolet - soft X-ray" (XUV) excess in LMC\,X-1. Any attempt to match the hard X-ray spectrum ($>1$ keV) with a conventional model of the accretion disk fails to account for this XUV component. We propose two likely sources for the observed anisotropy: (1) obscuration by a dusty torus, or (2) a jet-blown hole in a surrounding envelope of circumstellar absorbing material. We discuss the implications of our discovery in the context of the mass-scaling hypothesis for accretion onto black holes and suggest avenues for future research.


\end{abstract}


\section{Introduction}

The search for a so-called `unified model' of AGN has spanned more than two decades \citep{ant93}. Indeed, it has been shown that some Type 2 Seyferts may comprise a Type 1 Seyfert nucleus that is heavily obscured by a dense, dusty torus (\citealt{law82, ant85}). Although dusty tori obscure some of the nuclear continuum emission, a significant fraction escapes along the poles of the central source, producing a cone-shaped extended emission line region. Perhaps the most spectacular example occurs in the Type 2 Seyfert galaxy NGC\,5252 where bipolar \Oiii\ cones are seen to extend to $\sim$30 kpc in radius \citep{tad89}. This important phenomenon allows one to test the unified model by analyzing the ionization properties of the gas within the cones.

Many of the phenomena we observe from active galaxies arise from the properties of the accretion disk. \citet{ede99} note that the power density spectrum of an active galaxy resembles those observed in some Galactic black hole X-ray binaries (XRBs), such as Cyg\,X-1 (see also \citealt{mch89}). The transient behaviour of XRBs, however, may render `the scaling from stellar-to-supermassive black hole mass' a false argument \citep{don05}. But if appropriate corrections are made for the variations in the accretion rate, the variability timescales may be related \citep{mch06}.

If the mass-scaling hypothesis for black holes is correct, it is natural to question why we have not observed ionization cones associated with XRBs in previous work. This may reflect the difficulty of interpreting the complex nebulosity in the vicinity of XRBs due to a background of competing sources, e.g. hot young stars. The XRB LMC\,X-1 is surrounded by a complex filamentary emission-line nebula, catalogued by \citet{hen56} as N159F. The X-ray source ($L_{X}[2-10\,{\rm keV}] \approx2\times10^{38}\, {\rm erg\,\,s}^{-1}$, \citealt{sch94}) comprises a stellar mass ($4-10 \,M_\odot$) black hole accreting matter from an O7\,III-type star (``Star 32''). Based on the \Heii$\lambda4686$ emission line, \citet{pak86} demonstrated that N159F was the first example of an X-ray ionized nebula.  \citet{ram06} supported the presence of an X-ray ionized nebula, and concluded that LMC\,X-1 `is not injecting a significant amount of mechanical energy into the interstellar medium,' contrary to the findings of \citet{coo07} (hereafter, \papone), who recently suggested that the nebula surrounding LMC\,X-1 is largely driven by a jet emanating from the XRB. The latter study used integral field spectroscopy (IFS) to delineate the complex ionization regions within the nebula. The two studies prior to \papone\ used 1D slit spectroscopy, a technique that is spatially more limited than IFS.

Once again we appeal to IFS to obtain a complete spatial coverage of the nebula. The motivation behind our new study is to assess the extent of hard ionizing photons by tracing the \Heii\ and \Oiii\ emission. In the course of our work, we have discovered an ionization cone associated with this XRB, the first of its kind. Moreover, the ionization cone aligns with the putative jet from \papone, a phenomenon also observed in AGNs \citep{ung87}. Not only does this provide a new ground for testing accretion disk phenomena, but it also strengthens the mass-scaling hypothesis, implying that accretion processes around stellar-mass black holes are similar to accretion processes around supermassive black holes.

We summarize the observations and data reduction procedures in \S\,2. In \S\,3, we derive the cone parameters and discuss the ionizing source. Our interpretation and conclusions are summarized in \S\,4.

\section{Observations}

After presenting our previous results on LMC\,X-1 (\papone), we were alerted to archival VIsible Multi-Object Spectrograph (VIMOS) IFS observations (Program ID: 076.C-0284(B)) of N159F, which were acquired in $1.4''-1.6''$ seeing on 2005, October 1, and 27. Four 900\,sec VIMOS exposures were taken in high-resolution mode, two using the blue grism (R$ \, \approx 2050, 4150-6200 \, {\rm \AA}$) and two using the orange grism (R$ \, \approx 2150, 5200-7600\, {\rm \AA}$). The spectral range includes the prominent optical emission lines \Oi, \Oiii, \Hei, \Heii, \Ariii, \Nii, \Sii, \Hb\ and \Ha, all of which are important diagnostics for delineating regions containing multiple ionizing sources, and estimating the electron density and temperature. The two exposures with each grism were offset east-west by $12.7''$ ($19$ pixels) and north-south by $3.4''$ ($5$ pixels).

The data were reduced using the reduction pipeline VIPGI\footnote{VIPGI - Vimos Interactive Pipeline Graphical Interface, obtained from \emph{http://cosmos.iasf-milano.inaf.it/pandora/}} \citep{sco05}. The reduced data were converted into 3D datacubes and combined using the same data manipulation routines implemented in \papone. At the spectral resolution of the VIMOS grisms, all lines were found to be well approximated by a single unresolved Gaussian profile.

\section{Results}

\subsection{Morphology \& ionization source}\label{section:ionizationcone}

\begin{figure}
\includegraphics[width=5cm]{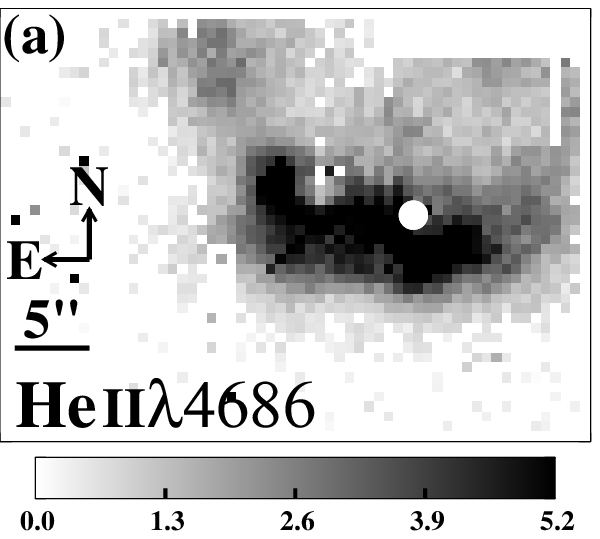}
\includegraphics[width=5cm]{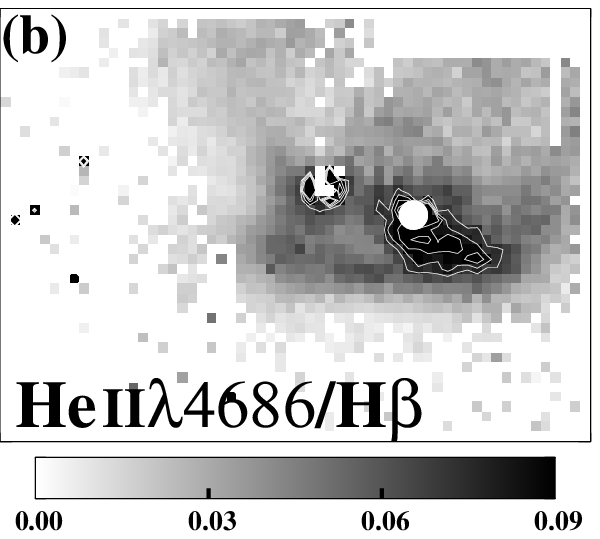}
\includegraphics[width=5cm]{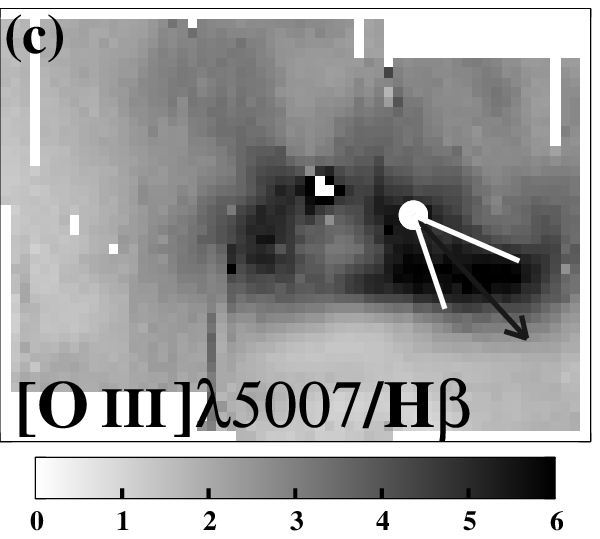}
\caption{In all figures the white dot corresponds to LMC\,X-1. (a) Enhanced \Heii$\lambda4686$ emission is observed to the south-east of LMC\,X-1. The emission is also present on extended scales, although it is much weaker. The units of the scalebar are $10^{-16}\,{\rm erg \, cm}^{-2}\,{\rm s}^{-1}\,{\rm arcsec}^{-2}$. (b) The \Heii$\lambda4686$/\Hb\ ratio highlights the ionization cone. Contours are overplotted with levels (0.07, 0.08, 0.09, 0.1), increasing towards the centre of the cone. (c) The ionization cone is also present in the \Oiii$\lambda5007$/\Hb\ ratio; the cone (two white lines extending from LMC\,X-1), and putative jet (gray arrow) from \papone\ are indicated.
\label{figure:HeII_OIII}}
\end{figure}

Fig.~\ref{figure:HeII_OIII}(a) immediately highlights the anisotropy of \Heii\ emission and its correlation with the \Oiii\ emission (see \papone), which is the first evidence that isotropic ionization by LMC\,X-1 is unlikely. An ionization cone is revealed when taking the ratio of these high ionization emission lines with respect to \Hb\ (see Fig.~\ref{figure:HeII_OIII}(b) \&\ (c)), as these line ratios are strongly dependent on the ratio of the ionizing photon intensity and the gas density (i.e. ionization parameter) at a fixed gas abundance. Jets, winds, and cones are often asymmetric and this may arise from the source or be due to large-scale dust or the presence of gas. The one-sided cone is observed in the direction of the putative jet (refer to \papone).

The \Oiii$\lambda\lambda(4959+5007)/4363$ ratio is a good diagnostic of nebular temperature, while the \Sii$\lambda\lambda6716/6731$ ratio provides an estimate of the electron number density when the ratio of this doublet is between $1.4$ ($\approx 30\, {\rm cm}^{-3}$) and $0.5$ ($\approx 10^{4}\, {\rm cm}^{-3}$) \citep{ost06}. We derive an electron temperature $T_e \approx 1 \times 10^4 \, {\rm K}$ and an electron number density $n_e \approx 75 \, {\rm cm}^{-3}$ in the ionization cone, with an uncertainty of about 10\%. The value of $n_e$ is revised downwards from \papone\ due to the improved data and is close to the low-density limit of this line diagnostic. It is noteworthy that outside the ionization cone the nebular temperature declines by $10-20\%$.

We note that LMC\,X-1 borders the giant \Hii\ region complex N159 \citep{hen56}. It is therefore likely that other nearby sources contribute to some of the ionization in N159F, which could explain some of the `excess' emission around LMC\,X-1. In \S3.2, we show that since the luminous stellar companion to LMC\,X-1 (Star 32) makes a negligible contribution to the cone energetics, the ionizing luminosity can only arise from LMC\,X-1. Furthermore, we find that the cone largely arises from photoionization rather than jet-powered shock ionization for two reasons. First, the nebula temperature is substantially cooler than expected for shock-heated gas at any metallicity \citep{dop03}. Secondly, radiative shock models predict an onset of \Heii$\lambda4686$ emission for shock speeds $\gtrsim 120 \, {\rm km} \, {\rm s}^{-1}$ \citep*{shu79,bin85,cox85,har87}. However, we showed in Paper I that the LMC\,X-1 jet is triggering a shock with a shock velocity that
is unlikely to exceed $v_s\,\approx 90\,{\rm km\,s}^{-1}$. 

\subsection{Cone parameters \& ionizing spectrum}\label{section:anisotropic}

Ionization cones provide a \emph{direct} measure of the flux of the ionizing continuum from the 
accretion disk if the geometry and orientation are clearly defined \citep*{mul96}. 
For the one-sided cone, we measure its position angle, half-opening angle and length to be PA$\,\simeq225\fdegs$, ${\cal H}_m \simeq 25\fdegs$ and $r_m\simeq12.5''\simeq3.3 \, {\rm pc}$ respectively; however, these are subject to the inclination angle of the cone to our line-of-sight, $i\sim60\fdegs$ \citep{cow92,mak00}. Therefore, the true half-opening angle, cone length and cone
solid angle are respectively ${\cal H}_c = \arctan[\tan{\cal H}_m\sin i] \simeq 22\fdegs$,
$r_c = r_m/\sin i\simeq3.8 \, {\rm pc}$, and $\Omega=2\pi(1-\cos{\cal H}_c)\simeq0.5 \,\, {\rm sr}$. For our inferred $n_e$ and $r_c$, the ionization cone is optically thick to H$^0$ Lyman continuum photons (${\cal N}({\rm H^0})$, $h\nu=13.6\,{\rm eV}$), and He$^+$ Lyman continuum photons (${\cal N}({\rm He^+})$, $h\nu=54.4\,{\rm eV}$). However, the ionization cone is optically thin to photons above $\epsilon_{u}=250-300\,{\rm eV}$ \citep*{yan98}.

We now use the observed X-ray spectrum to derive our first estimate of the ionizing flux from LMC\,X-1. The X-ray spectrum is best-fitted by a Comptonized multicolor disk (CMCD) model \citep*{yao05}. However, for energies $\lesssim1\,{\rm keV}$, the X-ray spectrum of LMC\,X-1 can be approximated by the simple multicolor disk (MCD) model (Fig.~2, \citealt{yao05}). The MCD model presented in Fig.~\ref{fig:xray_spectrum} adopts the following parameters: $kT_{in}=0.93\,{\rm keV}$ and $K_{\rm MCD}=57$, where $T_{in}$ is the inner disk temperature, and $K_{\rm MCD}$ is a normalizing constant. We present this model with and without line-of-sight attenuation ($N_H$).

Since $K_{\rm MCD}$ depends on the assumed inclination angle $i$, we set $i=0$ to determine the ionizing luminosity within the cone. We determine the absorption-corrected photon flux from $13.6\,{\rm eV}$ to $\epsilon_{u}=300\,{\rm eV}$ and $54.4\,{\rm eV}$ to $\epsilon_{u}=300\,{\rm eV}$ along our line-of-sight to be ${\cal F}_{\rm c}({\rm H^0})_{X}\approx0.6 \, {\rm phot} \, {\rm cm}^{-2} \, {\rm s}^{-1}$ and ${\cal F}_{\rm c}({\rm He^+})_{X}\approx0.4 \, {\rm phot} \, {\rm cm}^{-2} \, {\rm s}^{-1}$ respectively. Therefore the number of H$^0$ and He$^+$ Lyman continuum photons \emph{in the cone} produced by the X-ray source is
\begin{eqnarray}
{\cal N}_{\rm c}({\rm H^0})_X=\Omega D^2 {\cal F}_{\rm c}({\rm H^0})_{X} \simeq8\times10^{45} \, {\rm phot} \,\, {\rm s}^{-1}
\\
{\cal N}_{\rm c}({\rm He^+})_X=\Omega D^2 {\cal F}_{\rm c}({\rm He^+})_{X} \simeq6\times10^{45} \, {\rm phot} \,\, {\rm s}^{-1}
\end{eqnarray}
at the distance ($D\simeq55 \, {\rm kpc}$, \citealt{fea99}) to the LMC. The subscript `$X$' indicates that these were obtained from an extrapolation of the X-ray model.

From the VIMOS data, we measure the extinction corrected \Heii$\lambda4686$ flux of the cone to be $F_c(\lambda4686)\approx10^{-13} \, {\rm erg} \, {\rm cm}^{-2} \, {\rm s}^{-1}$ ($E$(B$-$V)$=0.37$, \citealt*{bia85}) corresponding to a \emph{cone} luminosity $L_c(\lambda4686)\approx4\times10^{34} \, {\rm erg} \,\, {\rm s}^{-1}$. If we assume case B recombination, from the ratio of the total to effective recombination coefficients, $\alpha({\rm He}^{+}, T)/\alpha_{4686}^{\rm eff}(T) = 4.2$ (for T=$10^4 \, {\rm K}$, \citealt*{peq91}), we estimate (Eq.~41, \citealt{har66})
\begin{equation}
{\cal N}_{\rm c}({\rm He^+}) =  \frac{L_c({\lambda 4686})}{h\nu_{4686}} \frac{\alpha({\rm He}^{+}, T)}{\alpha_{4686}^{\rm eff}(T)}      \simeq4\times10^{46} \, {\rm phot} \,\, {\rm s}^{-1}
\end{equation}
powers the cone, where $h\nu_{4686}$ $=$ 2.6 eV. This estimate would be higher still if the covering factor of the gas in the ionization cone was less than unity.

Considering the well-established metal deficiency of the LMC ($\sim\!\!\frac{1}{3}Z_{\odot}$), if Star 32 was \emph{solely} responsible for the ionization cone, its stellar temperature would need to be in excess of $50000 \, {\rm K}$ \citep{eva85}, which is inconsistent with the known value, $T_{\rm eff} = 37000 \, {\rm K}$ \citep{bia85}. However, Star 32 is capable of producing photons with energies $\gtrsim54.4\,{\rm eV}$, and assuming it radiates \emph{isotropically}, it will produce ${\cal N}({\rm He^+}) \simeq 6\times10^{44} \, {\rm phot} \,\, {\rm s}^{-1}$ \citep*{vac96}. Therefore Star 32 produces insufficient He$^+$ Lyman continuum photons to explain the observed ionization cone.

We now derive a new estimate of ${\cal N}_{\rm c}({\rm H^0})$ 
that is independent of the extrapolation from the X-ray spectrum. 
The \Oiii$\lambda5007$/\Oii$\lambda\lambda(3726+3729)$ ratio in the cone provides us with a direct 
determination of the ionization parameter $q$ for a gas with known metallicity \citep{kew02}.
Due to the insufficient spectral coverage of the VIMOS blue grism, the \Oii\ lines are not present in the VIMOS datacube. Instead, we use
the \Oiii/\Oii\ ratio ($\approx 1.9$) determined from ``Position 1" \citep{pak86}.
For the LMC gas phase abundance, this ratio corresponds to a unique ionization parameter $q_1\approx7\times10^{7} \, {\rm cm} \, {\rm s}^{-1}$ \citep{kew02} due to the differential dependence of oxygen ions on the ionization rate. Thus
\begin{equation}
{\cal N}_{\rm c}({\rm H^0})= \Omega \, r^2 \, q_1 \, n_e \simeq 8\times10^{46} \, {\rm phot} \,\, {\rm s}^{-1}
\end{equation}
where $r$ is the characteristic radius of the cone ($r_c/2$). Outside the cone, there is a decrease 
in $q$ at increasing radius from LMC\,X-1. For example, the \Oiii/\Oii\ ratio at a distance of $1.3'$ ($\sim20\,{\rm pc}$) north-east of LMC\,X-1 (``Position 2," \citealt{pak86}) is indicative of an ionization parameter $q_2\approx2\times10^{7} \, {\rm cm} \, {\rm s}^{-1}$. 

\section{Discussion}

\begin{figure}
\includegraphics[width=6.7cm,angle=-90]{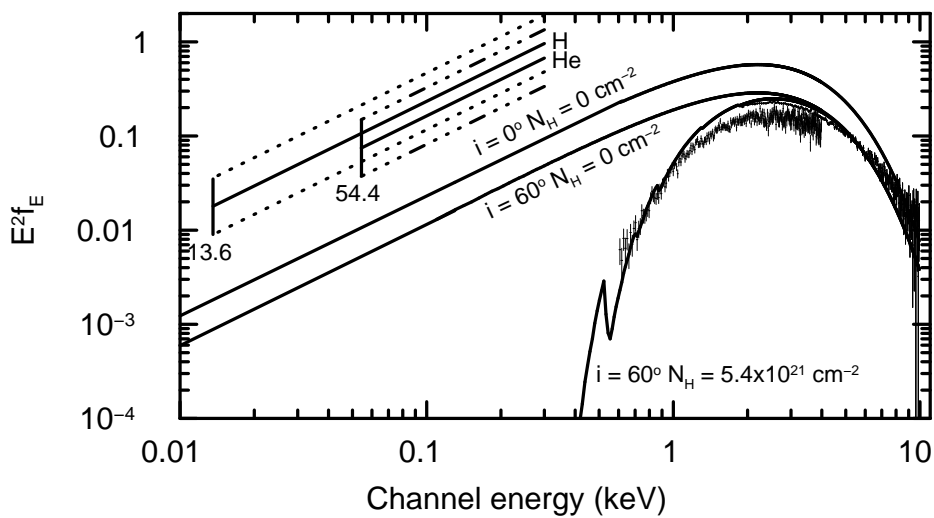}
\caption{The \emph{unfolded} X-ray spectrum of LMC\,X-1 (data points) with several models. The MCD approximation to the best fit CMCD model of LMC\,X-1 \citep{yao05} is presented with and without the line-of-sight attenuation, corresponding to the solid curves labelled ``$i=60\fdegs \,\, N_H=5.4\times10^{21}\,{\rm cm}^{-2}$'' and ``$i=60\fdegs \,\, N_H=0\,{\rm cm}^{-2}$'' respectively. According to this X-ray model, if we viewed LMC\,X-1 face-on (i.e. $i=0$) without the line-of-sight attenuation, we would observe the solid curve labelled ``$i=0\fdegs \,\, N_H=0\,{\rm cm}^{-2}$''. The solid curve labelled ``He'' corresponds to the \emph{direct} measure of the accretion disk XUV continuum (with $i=0\fdegs$ and $N_H=0\,{\rm cm}^{-2}$) in the energy range $54.4\,{\rm eV}$ to $300\,{\rm eV}$, derived from the \Heii$\lambda4686$ emission in the ionization cone, where the uncertainty in this curve is given by the two dot-dash curves. The solid curve labelled ``H'' corresponds to the accretion disk XUV continuum (with $i=0\fdegs$ and $N_H=0\,{\rm cm}^{-2}$) in the energy range $13.6\,{\rm eV}$ to $300\,{\rm eV}$, derived from the ionization parameter, where the uncertainty in this curve is given by the two dotted curves. The two solid vertical lines labelled $13.6\,{\rm eV}$ and $54.4\,{\rm eV}$ mark the ionization potentials of H and He respectively.}
\label{fig:xray_spectrum}
\end{figure}

By comparing Eq. (1)-(4), the Yao model underestimates the number of ionizing photons emerging from the accretion disk by a factor of $7-10$.  This discrepancy requires an extra component that dominates in the UV and/or soft X-ray bands.  In the Yao model, we cannot increase $K_{\rm MCD}$ because it is constrained beyond $1\,{\rm keV}$ by the observed hard X-ray spectrum (see Fig.~\ref{fig:xray_spectrum}), which is modelled as Comptonized disk emission.

We note that the size of the discrepancy arising from Eq.~(1)-(4) is exaggerated by the use of a photon number; the energy requirement is only a factor of two more than the Yao model. This could conceivably arise from upscattering of lower energy photons from more distant regions in the disk by a hot coronal plasma in the inner region. Alternatively, a sizeable fraction of the hard X-ray photons could be degraded to lower energies, i.e. in the opposite sense to what is assumed in the CMCD model. It is noteworthy that some AGN ionization cones can only be explained by the contribution of both ionization from the non-stellar nuclear continuum and jet-induced shock ionization \citep*{pog88,wil88}. Therefore an {\it in situ} jet may also contribute to the LMC\,X-1 cone, although the nebular temperature limits any contribution from shock processes.

EUV/soft X-ray excesses are not uncommon to AGN. The spectra of high redshift quasars, which can be observed in the UV, clearly exhibit a big blue bump component, widely attributed to accretion disk emission \citep{san89}. This component has also been inferred from the emission-line diagnostics of Seyfert 2 ionization cones \citep{ale00}. In some cases, EUV/soft X-ray excesses are \emph{also} observed. The origin of this component is unclear, but it may arise, for example, from thermal emission in a warm/hot cloud at a temperature that peaks in the $30-300\,$eV range \citep{sie95}. How this component relates to the accretion disk is not clear.

It is unlikely that the angular extent of the ionization cone reflects the poloidal ionizing field of the naked accretion disk. There are two plausible explanations for the anisotropic radiation pattern: (1) obscuration by a dusty torus, or (2) a jet-blown hole in the envelope of circumstellar absorbing material \citep*{kuu05,nes08}.

We suspect that obscuring XRB tori would scale to AGN tori with the mass accretion rate rather than the black hole mass. The mass accretion rate in XRBs depends on the donor star, whereas for AGN, it depends on the gas fuel supply from the surrounding interstellar medium. Because Star 32 is a giant, we would expect an excess mass flux to accumulate somewhere before reaching the accretion disk. This accumulated mass flux could form something akin to an obscuring torus. By analogy with AGN \citep{pie93}, high angular resolution, mid-infrared observations of LMC\,X-1 are required to confirm the presence of a dusty torus.

Here we have demonstrated the importance of ionization cones for establishing the XUV properties of obscured accretion disks. We suspect the ionization cone will also be observable in \Neiii$\lambda3868$ and \Nev$\lambda3426$ (ionization potential $63.5 \, {\rm eV}$ and $126.2 \, {\rm eV}$ respectively). These arise at higher $q$ values and will therefore allow us to probe down to smaller radii. Indeed, \citet{pak86} have already detected these two high ionization species in the immediate vicinity of LMC\,X-1. But due to the limited spatial coverage of their spectroscopic technique, we are unable to confirm that these ratios are enhanced in the direction of the ionization cone. Planned future observations will probe the structure of the ionization cone, allowing us to refine our ionization model and place tighter constraints on the nature of the ionizing source.

\acknowledgments

This discovery paper is dedicated to the memory of A.~S. Wilson at the University of Maryland who made seminal contributions to the field of active galactic nuclei. We thank an anonymous referee whose comments and suggestions improved the paper considerably. We also wish to thank R.~Soria for valuable discussions about {\sc xspec}, and for his assistance with X-ray spectral fitting. JBH is supported by a Federation Fellowship from the Australian Research Council.

{\it Facilities:} \facility{VLT (VIMOS)}.

\clearpage

\end{document}